\documentclass[10pt,twocolumn,aps,pra,showpacs]{revtex4}

\usepackage{bbm}

\begin{document}

\title{Optimal quantum searching to find a common element of two sets}         
\author{Avatar Tulsi\\
        {\small Department of Physics, IIT Bombay, Mumbai-400076, India}}  

\email{tulsi9@gmail.com}

\begin{abstract}
Given two sets $A$ and $B$ and two oracles $O_{A}$ and $O_{B}$ that can identify the elements of these sets respectively, the goal is to find an element common to both sets using minimum number of oracle queries. Each application of either $O_{A}$ or $O_{B}$ is taken as a single oracle query. This is basically a search problem and a straightforward application of Grover's algorithm can solve this problem but its performance is slow compared to the optimal one by a constant factor of 1.57. Here we present a variant of Grover's algorithm which achieves the optimal performance in not too restrictive cases. 
\end{abstract}


\maketitle

\section{Introduction}

In a general setting, Grover's algorithm~\cite{grover} drives a quantum system from a known initial state $|s\rangle$ to an unknown final state $|t\rangle$, which are also called \emph{source} and \emph{target} states respectively. This is done by iterating the operator $G = I_{s}I_{t}$ on $|s\rangle$, where $I_{\psi} = \mathbbm{1}-2|\psi\rangle \langle \psi|$ is the selective phase-inversion of the $|\psi\rangle$ state. It is known that $\pi/4|\langle s|t\rangle|$ iterations of $G$ on $|s\rangle$ will bring the quantum system very close to the desired final state $|t\rangle$. Though the operator $I_{s}$ is implemented using our knowledge of the source state $|s\rangle$, same is not true for $I_{t}$ which has to be implemented using an oracle $O$ that can identify and selectively apply a phase-inversion of the target state $|t\rangle$. Grover's algorithm is proved to be optimal~\cite{optimal}. 

In the case of search problem, we have a set of $N$ items labeled by $i \in \{0,1,\ldots, N-1\}$. Each item corresponds to a basis state $|i\rangle$ of a quantum system having an $N$-dimensional Hilbert space. We denote the set of solutions by $t$ and let $M = |t|$ be the total number of solutions. The source state is chosen to be the uniform superposition of all basis states $|s\rangle = (1/\sqrt{N})\sum_{i}|i\rangle$. The target state $|t\rangle$ is a uniform superposition of all target states $|t\rangle = (1/\sqrt{M})\sum_{i\in t} |i\rangle$. So, $|\langle s|t\rangle| = \sqrt{M/N}$ and $\frac{\pi}{4} \sqrt{N/M}$ iterations of $G$ are sufficient to get the state $|t\rangle$ which always gives a target state $i \in t$ upon measurement. To implement $I_{t}$, we define a binary function $f(i)$ which is $1$ for $i \in t$ else $0$. We attach an ancilla qubit, initially in state $|a\rangle$, to our search-space and then we use the oracle $O$ to compute $f(i)$ to get the transformation
\begin{equation}
|i\rangle |a\rangle \stackrel{O}{\longrightarrow} |i\rangle |a \oplus f(i)\rangle.  \label{groveroracle}
\end{equation}
Choosing $|a\rangle = |-\rangle = (|0\rangle - |1\rangle)/\sqrt{2}$, above transformation is equivalent to the $I_{t}$ transformation on the search space, i.e. $|i\rangle \stackrel{I_{t}}{\longrightarrow} (-1)^{f(i)}|i\rangle$.

We consider the problem where we are given two sets $A$ and $B$ and two oracles $O_{A}$ and $O_{B}$ which can apply selective phase-inversions on the elements of sets $A$ and $B$ respectively. For this problem, it is natural to consider each application of either $O_{A}$ or $O_{B}$ as a single oracle query. Our goal is to find an element $ i \in (A \cap B)$ which is common to both sets using minimum number of oracle queries. A quite related problem is the \emph{scheduling} problem, where two parties Alice and Bob has a list of calendar dates on which they are free. They want to schedule their meeting by finding a date on which both of them are free. They want to do so by using least amount of communication. This problem was first considerd by Buhrman, Cleve and Wigderson (BCW), where they had shown that qantum mechanics can give a quadratic improvement in the communication complexity of the problem~\cite{schedule1}. Later, Grover came up with an algorithm with a better communication complexity~\cite{schedule2}. 

Our problem is a simple variant of search problem and we can use Grover's algorithm to solve this problem by implementing the $I_{t}$ operator using the given oracles $O_{A}$ and $O_{B}$. As we show in next section, Grover's algorithm takes $\frac{\pi^{2}}{8}\sqrt{N/M}$ oracle queries to solve this problem where $M = |A \cap B|$ is the total number of elements common to both sets. 
The optimality proof of Grover's algorithm can be used to prove that any algorithm will take at least $q_{\rm opt} = \pi/4(\sqrt{N/M})$ queries. Because if an algorithm takes less than $q_{\rm opt}$ queries then by choosing $A = B$, we can get a faster quantum search algorithm which is impossible. Since Grover's algorithm takes $\frac{\pi^{2}}{8}\sqrt{N/M} = \frac{\pi}{2}q_{\rm opt} \approx 1.57 q_{\rm opt}$ queries for our problem, its performance is suboptimal by a constant factor. In Sec. III, we present a variant of Grover's algorithm which achieves the optimal performance in not too restrictive cases and then we conclude in Sec. IV.

\section{Using Grover's algorithm}

Here we discuss straightforward applications of Grover's algorithm to our problem. We present two such algorithms based on different approaches. These algorithms are motivated by the quantum algorithms for scheduling problem~\cite{schedule1,schedule2}.

\emph{Algorithm-1} In our problem, $t \in (A \cap B)$ and we need to implement $I_{t}$ using given oracles $O_{A}$ and $O_{B}$. Let us define two binary functions: $f_{A}(i)$ which is $1$ for $i \in A$ else $0$ and similarly, $f_{B}(i)$ which is $1$ for $i \in B$ else $0$. These functions can be computed using our oracles and we have 
\begin{equation}
|i\rangle |a\rangle \stackrel{O_{A}}{\longrightarrow} |i\rangle |a \oplus f_{A}(i)\rangle\ ,\ \ \ |i\rangle |a\rangle \stackrel{O_{B}}{\longrightarrow} |i\rangle |a \oplus f_{B}(i)\rangle. \label{twooracles}
\end{equation}
One possible way to implement $I_{t}$ is to attach an ancilla qubit with the initial state $|a\rangle = |0\rangle$. Then we use oracle $O_{A}$ and apply Hadamard gate $H_{anc}$ on the ancilla qubit to get 
\begin{equation}
|i\rangle|0\rangle \stackrel{O_{A}}{\longrightarrow} |i\rangle|f_{A}(i)\rangle \stackrel{H_{anc}}{\longrightarrow} |i\rangle\frac{|0\rangle +(-1)^{f_{A}(i)}|1\rangle}{\sqrt{2}}. \label{Aoracle}
\end{equation}   
The final state of ancilla qubit is $|-\rangle$ if $f_{A}(i) = 1$ otherwise $|+\rangle = (|0\rangle +|1\rangle)/\sqrt{2}$. Next, we use the oracle $O_{B}$. Since $|+ \oplus f_{B}(i)\rangle = 1^{f_{B}(i)}|+\rangle$ and $|- \oplus f_{B}(i)\rangle = (-1)^{f_{B}(i)}|-\rangle$, we find that the basis state $|i\rangle$ gets a phase factor of $-1$ iff $f_{A}(i) = f_{B}(i) = 1$, i.e. if $i \in (A \cap B)$ is a target state. Then the ancilla qubit can be decoupled from our system by doing the reverse of Eq. (\ref{Aoracle}) and effectively we implement $I_{t}$ on the search-space. This implementation uses $3$ oracle queries, $2$ of $O_{A}$ and $1$ of $O_{B}$, and since Grover's algorithm uses $q_{\rm opt} = \frac{\pi}{4}\sqrt{N/M}$ applications of $I_{t}$, we find that a total of $3q_{\rm opt}$ queries are needed to solve our problem using Algorithm-1.

\emph{Algorithm-2} 
A second approach  is to first drive our quantum system from the uniform superposition of all basis states $|s\rangle = (1/\sqrt{N})\sum_{i}|i\rangle$ to the uniform superposition of only those states corresponding to the elements of set $A$, i.e. $|s_{A}\rangle = \frac{1}{\sqrt{|A|}}\sum_{i \in A}|i\rangle$. This is done by treating $i \in A$ states as effective target states in Grover's algorithm.The corresponding $I_{t}$ transformation is then $I_{A}$, a selective phase inversion of $i \in A$ states, and it is implemented using the oracle $O_{A}$ that computes $f_{A}(i)$ in the same way as $I_{t}$ was implemented in Eq. (\ref{groveroracle}). We get the state $|s_{A}\rangle$ using $q_{A} = \pi/4(\sqrt{N/|A|})$ iterations of the operator $I_{s}I_{A}$ on $|s\rangle$. So we have 
\begin{equation}
 |s_{A}\rangle = \frac{1}{\sqrt{|A|}}\sum_{i \in A}|i\rangle = (I_{s}I_{A})^{q_{A}}|s\rangle.
\end{equation} 
Now let us consider $|s_{A}\rangle$ state as an effective source state and iterate the operator $I_{s_{A}}I_{B}$ on $|s_{A}\rangle$, where $I_{B}$ is the selective phase-inversion of $i \in B$ states, implemented using oracle $O_{B}$, and the operator $I_{s_{A}}$ is given by
\begin{eqnarray}
I_{s_{A}} & = & I-2|s_{A}\rangle \langle s_{A}| \nonumber \\
& =& I-2(I_{s}I_{A})^{q_{A}}|s\rangle\langle s|(I_{A}I_{s})^{q_{A}} \nonumber \\
& =& (I_{s}I_{A})^{q_{A}}I_{s}(I_{A}I_{s})^{q_{A}}.
\end{eqnarray}
Since the state $|s_{A}\rangle$ has close to zero amplitude in $i \not\in A$ states and $I_{B}$ inverts the phases of only $i\in B$ states, we find that $|\langle s_{A}|t\rangle| = |A\cap B|/|A|$, where $|t\rangle = (1/\sqrt{|A\cap B|})\sum_{i \in (A\cap B)}|i\rangle$ is the target state. Iterating $I_{s_{A}}I_{B}$ will amplify the amplitudes of only desired states and the number of iterations required to get $|t\rangle$ is $\pi/4(\sqrt{|A|/|A\cap B|})$. Above equation implies that each iteration of $I_{s_{A}}I_{B}$ requires $2q_{A}+1$ oracle queries, $2q_{A}$ for $I_{s_{A}}$ and $1$ for $I_{B}$. Hence the total number of oracle queries required by Algorithm-2 is 
\begin{equation}
\frac{\pi}{4}\sqrt{\frac{|A|}{|A\cap B|}}\left(2\frac{\pi}{4}\sqrt{\frac{N}{|A|}}+1\right) = \frac{\pi^{2}}{8}\sqrt{\frac{N}{|A\cap B|}} = \frac{\pi}{2} q_{\rm opt}\ . 
\end{equation}
Though this algorithm is better than Algorithm-1, this is slow compared to the optimal performance by a factor of $\pi/2 \approx 1.57$. So far, we are trying to use Grover's algorithm to solve our problem. In the next section, we present a variant of Grover's algorithm which naturally relates to our problem and solves it optimally.   

\section{Variant of Grover's algorithm}

\subsection{The search operator}
Here we present a variant of Grover's algorithm for our problem. Our algorithm starts with the uniform superposition of all basis states $|s\rangle = (1/\sqrt{N})\sum_{i}|i\rangle$ and then iterates the following operator on it.
\begin{equation}
V = I_{s}I_{B}I_{s}I_{A}. \label{newoperator}
\end{equation}
To analyse the iteration of $V$ on $|s\rangle$, we need to find its eigenspectra. We divide the set $N$ of all basis states $i \in \{0,1,\ldots,N-1\}$ in four mutually non-intersecting sets: (1) Set $A'$ whose elements are in set $A$ but not in set $B$, i.e. $A' = A - (A\cap B)$, (2) Set $B'$ whose elements are in set $B$ but not in set $A$, i.e. $B' = B- (A\cap B)$, (3) Set $T$ whose desired elements are common to both $A$ and $B$, i.e. $T = A\cap B$, and (4) Set $P$ whose elements are neither in set $A$, nor in set $B$, i.e. $P = N-(A \cup B)$. Correspondingly, we define four quantum states $|t\rangle$, $|p\rangle$, $|a\rangle$, and $|b\rangle$, which are the uniform superposition of basis states of the corresponding sets, i.e.
\begin{eqnarray}
|a\rangle = |A'|^{-1/2}\sum_{i \in A'}|i\rangle, &\ &  |b\rangle = |B'|^{-1/2}\sum_{i \in B'}|i\rangle,  \nonumber \\
|t\rangle = |T|^{-1/2}\sum_{i \in T}|i\rangle, & \ & |p\rangle = |P|^{-1/2}\sum_{i \in P}|i\rangle.
\end{eqnarray}
These states are mutually orthogonal to each other and the initial state $|s\rangle$ can be written in terms of them as
\begin{equation}
|s\rangle = \alpha|a\rangle + \beta |b\rangle + \gamma |t\rangle +\delta |p\rangle,\ \ \ \alpha^{2}+\beta^{2}+\gamma^{2}+\delta^{2} = 1.   \label{s4state}   
\end{equation}
where
\begin{equation}
\alpha = \sqrt{\frac{|T|}{N}},\ \beta = \sqrt{\frac{|P|}{N}},\ \gamma = \sqrt{\frac{|A'|}{N}},\ \delta = \sqrt{\frac{|B'|}{N}}.
\end{equation}

It is easy to see that our search operator $V$ preserves the four-dimensional subspace of $N$-dimensional Hilbert space. This subspace is orthogonally spanned by the states $|a\rangle$, $|b\rangle$, $|t\rangle$, and $|p\rangle$. The operator $I_{A}$ inverts the phases of $|a\rangle$ and $|t\rangle$ and the operator $I_{B}$ inverts the phases of $|b\rangle$ and $|t\rangle$. The $4 \times 4$ matrix representation of these operators in this subspace are given by
\begin{equation}
I_{A} = \left[ \begin{array}{cccc} -1 & 0 & 0 & 0 \\ 0 & 1 & 0 & 0 \\ 0 & 0 & -1 & 0 \\ 0 & 0 & 0 & 1 \end{array} \right],\ \ \ \ \  I_{B} = \left[ \begin{array}{cccc} 1 & 0 & 0 & 0 \\ 0 & -1 & 0 & 0 \\ 0 & 0 & -1 & 0 \\ 0 & 0 & 0 & 1 \end{array} \right]\ .
\end{equation}
To find the matrix representation of $I_{s}$, we use the Eq. (\ref{s4state}) and the expression $I_{s} = I-2|s\rangle\langle s|$. We find it to be
\begin{equation}
I_{s} = \left[\begin{array}{cccc} 1-2\alpha^{2} & -2\alpha \beta & -2\alpha \gamma & -2\alpha \delta \\ -2\beta \alpha & 1- 2\beta^{2}& -2\beta\gamma & -2\beta \delta \\ -2\gamma \alpha & -2 \gamma \beta & 1- 2\gamma^{2} & -2\gamma \delta \\ -2\delta\alpha & -2\delta \beta & -2\delta \gamma & 1-2\delta^{2} \end{array} \right] \ .
\end{equation}
Putting these expression in Eq. (\ref{newoperator}), we find that the matrix representation of our search operator $V$ in the preserved four-dimensional subspace is given by
\begin{equation}
V = \left[\begin{array}{cccc} 8\alpha^{2}r-1 & 4\alpha \beta l_{2} & -4\alpha \gamma l_{2} & -8\alpha \delta r \\ -4\alpha \beta l_{2} & 8\beta^{2} l_{1} - 1 & -8\beta \gamma l_{1} & 4\beta \delta l_{2} \\ -4\alpha\gamma l_{2} & 8\beta\gamma l_{1} & 1- 8\gamma^{2} l_{1} & 4\gamma\delta l_{2} \\ 8\alpha \delta r & 4\beta \delta l_{2} & -4\gamma\delta l_{2} & 1-8\delta^{2}r  \end{array}\right]\ ,
\end{equation}
where $r = \beta^{2} +\gamma^{2}$ and $l_{k} = 1-kr$. Next, we find the eigenspectra of $V$.


\subsection{Eigenvalues}
To find the eigenvalues of $V$, we note that this is a real operator, so its eigenvalues come in complex conjugate pairs. Further, we make an important observation that
\begin{equation}
 V^{T}_{ij} = (-1)^{i+j}V_{ij}\  \Longrightarrow\  V^{\dagger}_{ij} = (-1)^{i+j}V_{ij}\ .  \label{Vtranspose}
\end{equation} 
The latter identity is because $V$ is a real unitary operator, so $V^{T} = V^{\dagger}$. Let $|\psi\rangle$ be an eigenstate of $V$ with eigenvalue $e^{\imath \theta}$, i.e. $V|\psi\rangle = e^{\imath \theta}|\psi\rangle$. Then 
\begin{eqnarray}
(V + V^{\dagger})|\psi\rangle & = & (2\cos\theta)|\psi\rangle \\
(V - V^{\dagger})|\psi\rangle & = & (-2\imath \sin \theta)|\psi\rangle .
\end{eqnarray}
Using Eq. (\ref{Vtranspose}), we get
\begin{eqnarray}
\sum_{j} [1+(-1)^{i+j}] V_{ij}\psi_{j} = 2\cos\theta \psi_{j}  \nonumber \\  
\sum_{j} [1- (-1)^{i+j}] V_{ij}\psi_{j} = 2 \imath \sin \theta \psi_{j}\ .   \label{realimgn}
\end{eqnarray}
First of the above two equations can be written as eigenvalue equations of two $2\times 2$ sub-matrices, $V_{\rm odd}$ and $V_{\rm even}$, of $V$ as 
\begin{equation}
V_{\rm odd}\left[\begin{array}{c} \psi_{1} \\ \psi_{3}  \end{array}\right]  =    \left[\begin{array}{cc} V_{11} & V_{13} \\ V_{13} & V_{33} \end{array}\right]\left[\begin{array}{c} \psi_{1} \\ \psi_{3}  \end{array}\right] = \cos\theta \left[\begin{array}{c} \psi_{1} \\ \psi_{3}  \end{array}\right]   \label{Voddeqn}
\end{equation}
and
\begin{equation}
V_{\rm even}\left[\begin{array}{c} \psi_{2} \\ \psi_{4}  \end{array}\right] = \left[\begin{array}{cc} V_{22} & V_{24} \\ V_{24} & V_{44} \end{array}\right]\left[\begin{array}{c} \psi_{2} \\ \psi_{4}  \end{array}\right] = \cos\theta \left[\begin{array}{c} \psi_{2} \\ \psi_{4}  \end{array}\right]  \label{Veveneqn}
\end{equation}
It is easy to verify that the two eigenvalues, $\cos \theta _{\pm}$, of sub-matrices $V_{\rm odd}$ and $V_{\rm even}$ are same and they determine the four eigenvalues $e^{ \pm \imath \theta_{\rm \pm}}$ of $V$. We have
\begin{eqnarray}
\cos \theta_{\pm} &=& \chi \pm \sqrt{\zeta + \eta} \nonumber \\
 \chi &=& (V_{11}+V_{33})/2  \nonumber \\
\zeta &=& (V_{11}-V_{33})^{2}/4 \nonumber \\
\eta & =& V_{13}^{2}\ .   \label{eigenvalues}
\end{eqnarray}
Using expressions for $\{V_{11},V_{13},V_{33}\}$, we get
\begin{eqnarray}
\chi &=& 4(\alpha^{2}+ \gamma^{2})(\beta^{2}+\gamma^{2}) - 4\gamma^{2}\ , \nonumber \\ 
\zeta &=& [1-4(\beta^{2}+\gamma^{2})(\alpha^{2}-\gamma^{2})-4\gamma^{2}]^{2}\ , \nonumber \\
\eta & = & [4\alpha \gamma (2\beta^{2} + 2\gamma^{2} - 1)]^{2}    \label{chizetaeta}
\end{eqnarray}
Next, we find the eigenvectors of $V$

\subsection{Eigenvectors}

The eigenvectors of $V$ are found by finding the eigenvectors of $V_{\rm odd}$ and $V_{\rm even}$. Eqs. (\ref{Voddeqn}) and (\ref{Veveneqn}) gives the following relationships for the eigenvectors
\begin{eqnarray}
\psi_{3} & = & g_{\pm}\psi_{1}\ , \ \ \ g_{\pm} = (\cos \theta_{\pm}-V_{11})/V_{13}\ , \nonumber \\ 
\psi_{4} & = & h_{\pm}\psi_{2}\ , \ \ \ h_{\pm} = (\cos \theta_{\pm}-V_{22})/V_{24} 
\end{eqnarray}
Since the off-diagonal elements of $V_{\rm odd}$ and $V_{\rm even}$ are equal, we have $g_{+}g_{-} = h_{+}h_{-} = -1$ so that we can write $g_{\pm} = \pm g^{\pm 1}$ and $h_{\pm} = \pm h^{\pm 1}$. Also, Eq. (\ref{realimgn}) implies that if we choose $\psi_{1}$ to be real, then $\psi_{3}$ is also real while $\psi_{2}$ and $\psi_{4}$ are imaginary. We also note that $V|\psi\rangle = e^{\imath \theta}|\psi\rangle$ implies that $V|\psi^{*}\rangle = e^{-\imath \theta}|\psi^{*}\rangle$ since $V$ is real. 

With all these observations, we can write down the diagonalizer matrix $D$ whose columns are the eigenvectors of $V$ i.e.,
\begin{equation}
D = \left[\begin{array}{cccc} x & x & gx & gx \\ \imath y & - \imath y & \imath hy & -\imath hy \\ gx & gx & -x & -x \\ \imath hy & -\imath hy & -\imath y & \imath y \end{array}\right],  \label{Dmatrix}
\end{equation}  
and we have
\begin{equation}
DVD^{\dagger} = \left[\begin{array}{cccc} e^{\imath \theta_{+}} & 0 & 0 &  0 \\ 0 & e^{-\imath \theta_{+}}& 0& 0 \\ 0 & 0& e^{\imath \theta_{-}}& 0 \\ 0 & 0 & 0 & e^{-\imath \theta_{-}} \end{array}\right].
\end{equation} 
Here $x$ and $y$ are the arbitrary constants in the choice of eigenvectors for $V_{\rm odd}$ and $V_{\rm even}$ respectively. To fix these constants, we use the normalization condition, i.e. choose the column vectors of $D$ to be of unit length. Note that the first two column vectors of $D$, $|D_{i1}\rangle$ and $|D_{i2}\rangle$, are mutually orthogonal and the complex conjugates of each other. If we add them, then the length of sum $|D_{i1}\rangle + |D_{i2}\rangle = 2(x|a\rangle -x/g |t\rangle)$ is $2$. Similarly, if we subtract them, then the length of the difference $|D_{i1}\rangle - |D_{i2}\rangle = 2\imath (y|a\rangle -y/h |t\rangle)$ is $2$. These give us
\begin{equation}
x = \frac{1}{\sqrt{2(1+g^{2})}}\ ; \ \ \ \ y = \frac{1}{\sqrt{2(1+h^{2})}}\ , \label{xy}
\end{equation}  
where 
\begin{eqnarray}
g = g_{+} &=& (\cos \theta_{+}-V_{11})/V_{13}\ ; \nonumber \\
 h = h_{+}& =& (\cos \theta_{+}-V_{22})/V_{24}\ . \label{gh}
\end{eqnarray}
These equations determine the eigenvectors of $V$. Next, we analyse the performance of our algorithm.  

\subsection{Performance}

To discuss the performance, we need to analyse the iteration of $V$ on the initial state $|s\rangle$ given by Eq. (\ref{s4state}). This can be done by working in the eigenbasis of $V$. To write $|s\rangle$ in this basis, we multiply $|s\rangle$ by $D^{\dagger}$ to get
\begin{equation}
|s\rangle = \sigma |\psi_{++}\rangle + \sigma^{*}|\psi_{-+}\rangle + \kappa |\psi_{+-}\rangle + \kappa^{*} |\psi_{--}\rangle\ ,
\end{equation}
where $|\psi_{\pm \pm}\rangle$ denote the normalised eigenvectors of $V$ with the eigenvalues $e^{\pm\imath \theta_{\pm}}$ (for example $|\psi_{+-}\rangle$ is the normalised eigenvector with eigenvalue $e^{+\imath \theta_{-}}$) and
\begin{eqnarray}
\sigma &=& x(\alpha + g\gamma) - \imath y (\beta + h\delta) \nonumber \\
\kappa &=& x(\alpha \delta - \gamma) - \imath y(h\beta - \delta)\ . \label{sigmakappa}
\end{eqnarray}
Thus we have
\begin{eqnarray}
V^{q}|s\rangle &=& \sigma e^{\imath q\theta_{+}}|\psi_{++}\rangle + \sigma^{*}e^{-\imath q\theta_{+}}|\psi_{-+}\rangle \\
 & +& \kappa e^{\imath q\theta_{-}}|\psi_{+-}\rangle + \kappa^{*} e^{-\imath q\theta_{-}}|\psi_{--}\rangle.
\end{eqnarray}
Now, Eq. (\ref{Dmatrix}) gives
\begin{equation}
|t\rangle = gx(|\psi_{++}\rangle + |\psi_{-+}\rangle) - x (|\psi_{+-}\rangle + |\psi_{--}\rangle)\ ,
\end{equation}
and hence
\begin{equation}
\langle t|V^{q}|s\rangle = 2gx|\sigma|\cos(\angle \sigma + q\theta_{+}) - 2x |\kappa| \cos (\angle \kappa + q\theta_{-}).  \label{performance}
\end{equation}
The probability of getting a target state after $q$ iterations of $V$ on the initial state is given by the square of above quantity. So far, we have presented the analysis of general case. Next, we consider the case when our algorithm achieves the optimal performance.

\subsection {Optimal Performance} 
We consider a special case when 
\begin{equation}
\gamma \ll 1,\ \ \ \ {\alpha,\beta} \leq 1/10 \label{optimalcase}   
\end{equation}
Note that $\gamma \ll 1$ is not at all a restrictive case as if $\gamma$ is comparable to $1$, then we can easily find the desired item by random trials. As $\gamma \ll 1$, we ignore $O(\gamma^{3})$ and higher terms in Eq. (\ref{chizetaeta}) to get
\begin{eqnarray}
\chi & = & 4\alpha^{2}\beta^{2} + 4\gamma^{2}(\alpha^{2}+\beta^{2}-1) \nonumber \\
\zeta & = & (1-4\alpha^{2}\beta^{2})^{2} + 8\gamma^{2}(\beta^{2} - \alpha^{2} -1)(1-4\alpha^{2}\beta^{2}) \nonumber \\
\eta & = & 16\gamma^{2}\alpha^{2}(2\beta^{2}-1)^{2}
\end{eqnarray}
After little calculation, we get
\begin{equation}
\sqrt{\zeta + \eta} =  (1-4\alpha^{2}\beta^{2}) + \gamma^{2}\left[4(\beta^{2}-\alpha^{2}-1) + \frac{8\alpha^{2}(2\beta^{2}-1)^{2}}{1-4\alpha^{2}\beta^{2}}\right].
\end{equation}
Using last two equations in Eq. (\ref{eigenvalues}), we get
\begin{equation}
\cos \theta_{+} = 1 - 8\gamma^{2}\frac{1-\alpha^{2}-\beta^{2}}{1-4\alpha^{2}\beta^{2}} \Longrightarrow \theta_{+} = 4\gamma \sqrt{\frac{1-\alpha^{2}-\beta^{2}}{1-4\alpha^{2}\beta^{2}}}\ , 
\end{equation}
and $\cos \theta_{-} = -1+8\alpha^{2}\beta^{2} + O(\gamma^{2})$. From now onwards, we ignore $O(\alpha,\beta,\gamma)$ terms in our calculations. Using Eq. (\ref{gh}), we get
\begin{equation}
g = -(2\alpha \gamma)^{-1} \gg 1;\ \ \ h = 1/2\beta.
\end{equation}
Then Eq. (\ref{xy}) gives
\begin{equation}
x = \sqrt{2}\alpha \gamma \ll 1;\ \ \ y = \sqrt{2}\beta\ .
\end{equation}
and Eq. (\ref{sigmakappa}) gives
\begin{equation}
\sigma = -\imath/\sqrt{2} + O(\beta^{2}) \Longrightarrow \angle\sigma \approx -\pi/2;\ \ \kappa =  \imath \beta/\sqrt{2} + O(\beta^{2})\ .
\end{equation}
Since $g \gg 1$ and $|\sigma| > |\kappa|$, we find that the second term in the R.H.S. of Eq. (\ref{performance}) can be neglected and using above equations, we get
\begin{equation}
|\langle t|V^{q}|s\rangle|  = \cos\left(\frac{\pi}{2} + 4q\gamma\right) + O(\beta,\alpha,\gamma).
\end{equation}
So we get a near to $1$ amplitude in the target state after $\pi /8\gamma$ iterations of our search operator $V$. Since each iteration of $V$ takes $2$ oracle queries, $1$ for $O_{A}$ and $1$ for $O_{B}$, the total number of queries used by our algorithm is $q_{\rm opt} = \pi/4\gamma$, which is also the optimal one. Thus, our algorithm is optimal as long as our problem parameters satisfy the condition (\ref{optimalcase}) which is, by no means, too restrictive.

\section{Conclusion}

We have presented a variant of quantum search algorithm, which naturally relates to our goal of finding a common element of two sets. We have shown that our algorithm can be exactly analysed by considering a four-dimensional subspace and by using the properties of the matrix representing our search operator. Though a straightforward application of standard quantum searching will give a sub-optimal performance, our algorithm can give optimal performances in not too restrictive conditions.


\begin{thebibliography}{99}
\bibitem{grover} L.K. Grover, Phys. Rev. Lett. \textbf{79}, 325 (1997). 
\bibitem{optimal} C. Bennett, E. Bernstein, G. Brassard, and U. Vazirani, SIAM J. Computing \textbf{26}, 1510 (1997).
\bibitem{schedule1} H. Buhrman, R. Cleve, and A. Wigderson, Proc. 30th STOC, 1998, 63--68.
\bibitem{schedule2} L.K. Grover, arXiv.org:quant-ph/0202033.
\end{thebibliography}
\end{document}